\documentclass[conference]{IEEEtran}
\IEEEoverridecommandlockouts
\usepackage{cite}
\usepackage{amsmath,amssymb,amsfonts}
\usepackage{algorithmic}
\usepackage{enumitem}
\usepackage{subfigure}
\usepackage{graphicx}
\usepackage{textcomp}
\usepackage{gensymb}
\usepackage{xcolor}
\usepackage{soul}

\usepackage{float}
\usepackage{amssymb}
\usepackage{amsfonts} 
\usepackage{graphicx} 
\usepackage{algorithm}
\usepackage[update,prepend]{epstopdf}
\usepackage{cite}
\usepackage{xcolor}
\usepackage{soul}
\usepackage{caption}
\usepackage{gensymb}
\usepackage{etoolbox}
\usepackage{rotating}
\usepackage{comment}
\usepackage{balance}
\usepackage{booktabs}
\usepackage{multirow}
\usepackage{hyperref}
\usepackage{CJK} 
\usepackage{tikz}

\def\BibTeX{{\rm B\kern-.05em{\sc i\kern-.025em b}\kern-.08em
    T\kern-.1667em\lower.7ex\hbox{E}\kern-.125emX}}
\begin{document}

\title{Hand Gesture Recognition through \\ Reflected Infrared Light Wave Signals\\
\thanks{This material is based upon work supported in part by the U.S. Department of Energy, Office of Science, Office of Advanced Scientific Computing Research under Award Number DE-SC0023023. This work was supported in part by the U.S. National Science Foundation under Grant 2008556.}

\thanks{This work has been submitted to the IEEE for possible publication.
Copyright may be transferred without notice, after which this version may
no longer be accessible.}
}

\author{\hspace{-.6cm}
\IEEEauthorblockN{Md Zobaer Islam}
\IEEEauthorblockA{\hspace{-.6cm}\textit{School of Electrical and Computer}\\
\hspace{-.6cm}\textit{Engineering}\\
\hspace{-.6cm}\textit{Oklahoma State University}\\
\hspace{-.6cm}Stillwater, OK, USA \\
\hspace{-.6cm}zobaer.islam@@okstate.edu}
\and
\hspace{-1cm}
\IEEEauthorblockN{Li Yu}
\IEEEauthorblockA{\hspace{-.6cm}\textit{Electrical and Computer Engineering}\\
\hspace{-.6cm}\textit{University of Windsor}\\
\hspace{-.6cm}Windsor, Ontario, Canada \\
\hspace{-.6cm}yuli@uwindsor.ca}
\and
\hspace{-.7cm}
\IEEEauthorblockN{Hisham Abuella}
\IEEEauthorblockA{\hspace{-.7cm}\textit{School of Electrical and Computer}\\
\hspace{-.7cm}\textit{Engineering}\\
\hspace{-.7cm}\textit{Oklahoma State University}\\
\hspace{-.7cm}Stillwater, OK, USA \\
\hspace{-.7cm}hisham.abuella@okstate.edu}
\and

\hspace{1cm}
\IEEEauthorblockN{John F. O'Hara}
\IEEEauthorblockA{\hspace{1cm}\textit{School of Electrical and Computer}\\
\textit{\hspace{1cm}Engineering}\\
\textit{\hspace{1cm}Oklahoma State University}\\
\hspace{1cm}Stillwater, OK, USA \\
\hspace{1cm}oharaj@okstate.edu}
\and
\hspace{.9cm}
\IEEEauthorblockN{Christopher Crick}
\IEEEauthorblockA{\hspace{1cm}\textit{Computer Science Department}\\
\textit{\hspace{1cm}Oklahoma State University}\\
\hspace{1cm}Stillwater, OK, USA \\
\hspace{1cm}chris.crick@okstate.edu}
\and
\hspace{.86cm}
\IEEEauthorblockN{Sabit Ekin}
\IEEEauthorblockA{\hspace{1cm}\textit{Department of Engineering Technology}\\
\textit{\hspace{1cm}and Industrial Distribution}\\
\textit{\hspace{1cm}Texas A\&M University}\\
\hspace{1cm}College Station, TX, USA \\
\hspace{1cm}sabitekin@tamu.edu}
}



\maketitle

\begin{abstract}
In this study, we present a wireless (non-contact) gesture recognition method using only incoherent light wave signals reflected from a human subject. In comparison to existing radar, light shadow, sound and camera-based sensing systems, this technology uses a low-cost ubiquitous light source (e.g., infrared LED) to send light towards the subject's hand performing gestures and the reflected light is collected by a light sensor (e.g., photodetector). This light wave sensing system recognizes different gestures from the variations of the received light intensity within a 20-35\,cm range. The hand gesture recognition results demonstrate up to 96\% accuracy on average. The developed  system can be utilized in numerous Human-computer Interaction (HCI) applications as a low-cost and non-contact gesture recognition technology.

\end{abstract}


\section{Introduction}
Recent trends in the computer and communication industries, Internet of Things (IoT) and the application of computers in medicine show that human-computer interaction (HCI) is becoming an increasingly important technological discipline. Research on HCI is crucial for creating complex, computerized systems that can be operated intuitively and efficiently by people without any formal training. Well-designed HCI interfaces make it convenient to control machines for education, labor, communication, healthcare, and entertainment environments\cite{IOT1,hcixin}.
Hand gesture recognition is a natural choice for HCI.  Simple movements of the hand can represent a type of sign language to machines resulting in the execution of complex actions. 

Existing hand gesture recognition techniques can be classified into two groups: wearable sensing and remote (non-contact) sensing. In wearable sensing, the user literally wears the sensor(s), which may be installed on a glove or otherwise attached to the hand~\cite{wear1,wear4}. While this sensing mode is both stable and responsive, the sensor(s) must be worn whenever hand movement is to be detected which is inconvenient. In remote sensing, hand gestures are perceived without any special hardware attached to the hand. The most frequently used sensors utilize radio frequency (RF) waves, light shadows, cameras, and sound waves~\cite{RF1,RF2,RF3,RF4, image1,image2, light7,light16, soundnew1,soundnew2}. RF-based sensing systems are prone to electromagnetic interference~\cite{emirf1,emirf2}. Also, RF signals can easily penetrate a
human body, and the transmitted RF power level becomes a
safety concern~\cite{geisheimer2001non, masao2001biological}, especially when multiple RF devices are present in the surrounding. The light-shadow based systems employ photodetectors in the ceiling or floor to capture the shadow patterns created by ambient light due to hand gestures, but interference from obstacles between the body and the receiver can be a critical issue here. In camera-based gesture recognition systems, the input data are images and/or videos which require enhanced processing power and large storage. Also, privacy issues associated to capturing human pictures must be taken into account~\cite{camera5,camera6}. The sound-based gesture recognition systems have good accuracy but the ultrasonic frequency they use may harm or perturb children and pets~\cite{soundnew}.

We have developed a hand gesture recognition system by utilizing reflected incoherent infrared light wave signals - a novel sensing method for hand gesture recognition termed as light wave sensing (LWS). The main functional components include the LWS hardware, the signal processing algorithms, and the classification algorithms for gesture detection.  Infrared LED (light-emitting diode) light sources are used to illuminate the hand performing the gesture. The reflected intensity of the light varies with the movement of the hand and is captured and converted into an electrical signal by a commercially-available photodetector.  The time-domain variations of the received (raw) signals are filtered by signal processing algorithms and then classified by using machine learning tools, which are trained by prior captured data sets. With this modality, we can distinguish hand gestures with up to 96\% accuracy with infrared sensing at 20\,cm sensing distance.


The rest of this paper is organized as follows. Section~\ref{sec:System_Design} describes the principles and system design, both in terms of hardware and software algorithms utilized.  Section~\ref{sec:Results} presents the  evaluation of the LWS system, and some brief related discussions. Finally, Section~\ref{sec:Conclusion} presents the conclusions and future work. 

\section{System Design and Implementation}
\label{sec:System_Design}
\subsection{Sensing Hardware}
\begin{figure}[!t]
\centering
\includegraphics[width=0.38\textwidth]{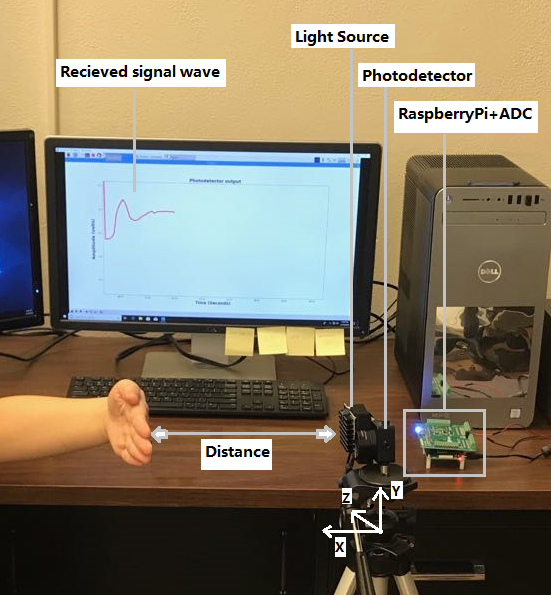}
\vspace{-1mm}
\caption{{The experimental setup of our light wave sensing  system.}}
\label{fig:Hardware_setup}
\vspace{-4mm}
\end{figure}

{Our LWS system setup consists of infrared light sources as transmitter, a photodetector as receiver, a digital signal processing (DSP) unit to convert the received signal from analog to digital, and an electronic module to process and store the digital data. The light source was invisible 940\,nm IR lamp board with light sensor (48 black LED illuminator array) having 30\,ft range and 120\degree~wide beamwidth~\cite{transmitter}. A commercial photodetector, Thorlabs PDA100A, was served as the receiver~\cite{detector}. A Raspberry Pi with a PiPlate analog to digital converter (ADC) circuit handled the data collection, digitization and storage for offline processing. The overall experimental setup is shown in Fig.~\ref{fig:Hardware_setup}.} 

\subsection{Measurement Procedure Overview}
{In the measurement operation, volunteers were asked to perform 
gestures with their hands in the designated area at a distance $d$ in front of the receiver (and also the transmitter). The infrared transmitter was targeted such that the hand was centered at the brightest part of the transmitted beam. An infrared monitor was used to optimize the aiming of the beam. The photodetector received the reflected light intensity waves from the hand of the volunteer which were unique according to the changing distance, shape and scattering cross-section of the hand. Light intensity data were recorded for 6 seconds at a sampling rate of 100\,Hz, though each gesture might be finished in 2-3 seconds. This resulted in individual gesture data sets of approximately 600 bytes (single precision). 
The data were then processed and classified offline using various signal processing and machine learning algorithms to be discussed in the next subsections.}

\begin{figure}[t]
\includegraphics[width=0.3\textwidth]{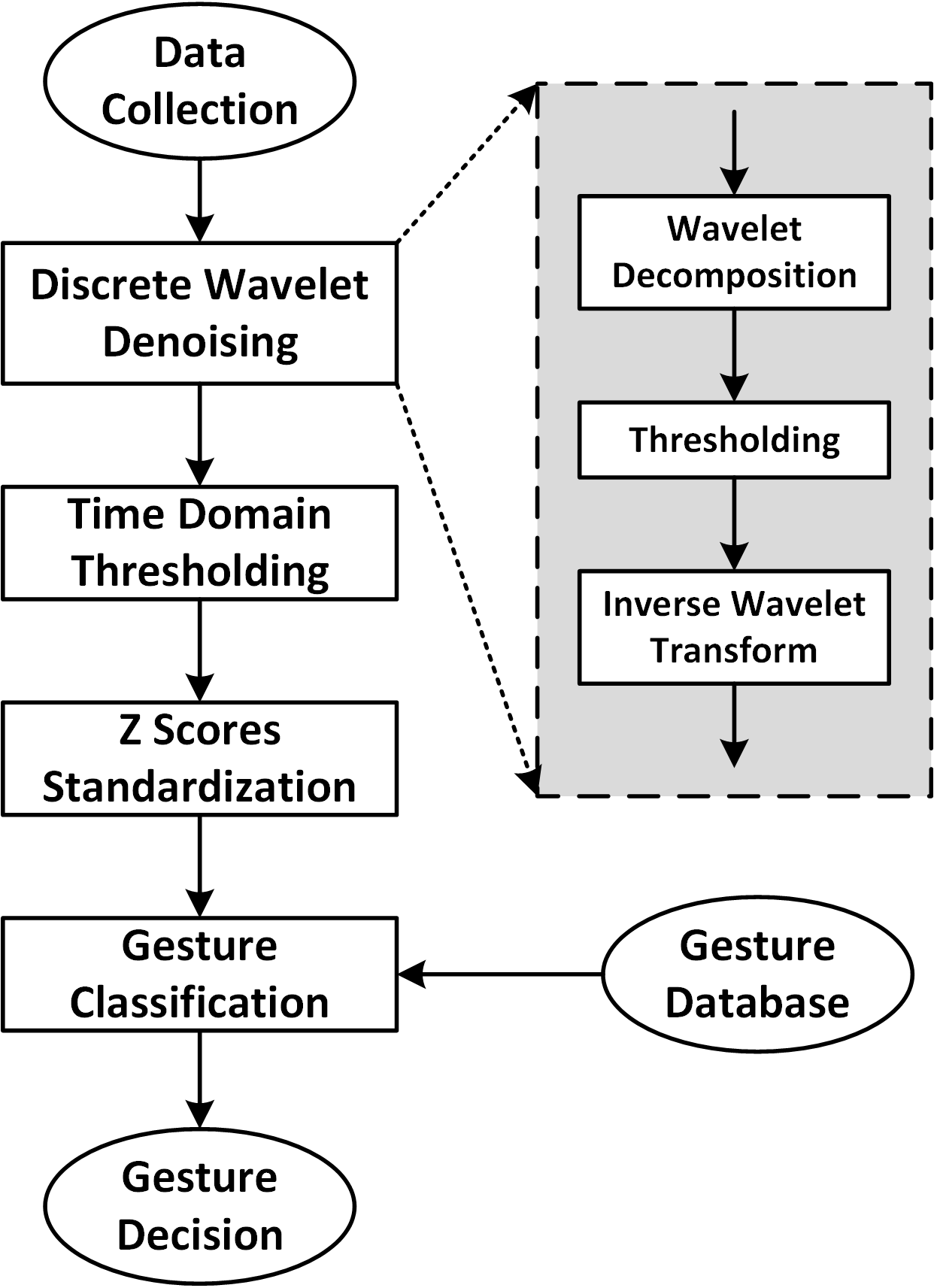} 
\centering
\caption{Flow diagram of the steps and algorithms used to prepare measured data for gesture classification.}
\label{fig:Flow_Diagram}
\vspace{-4mm}
\end{figure}

\subsection{Signal Processing}

In order to extract the patterns and important signal features resulting from different gestures, and to remove redundant information and noise from the received signal, multiple signal processing algorithms were needed. A flow diagram of the sequential operations performed on the raw data is presented in Fig.~\ref{fig:Flow_Diagram}. At first, the signal corrupted with noise and interference effects was denoised using Discrete Wavelet Denoising methods. Two noise features that were prominent in the frequency-domain signal appeared to be due to the flicker of ceiling lights, evident at 120\,Hz,  and the flicker of nearby computer monitors, evident at 60\,Hz. Discrete Wavelet Transform (DWT) is highly useful in analyzing non-stationary signals since it provides both time-domain and frequency-domain representation of the signal~\cite{dwt, dwtxin}. Prior to denoising, the signal was decomposed into wavelets using DWT. Then, the high frequency noise coefficients below an appropriately chosen threshold value in the decomposed signal were suppressed using the wavelet thresholding method. In the following, Inverse DWT used the non-zero coefficients to generate the denoised signal. 

\begin{figure}[htbp]
\includegraphics[width=0.44\textwidth]{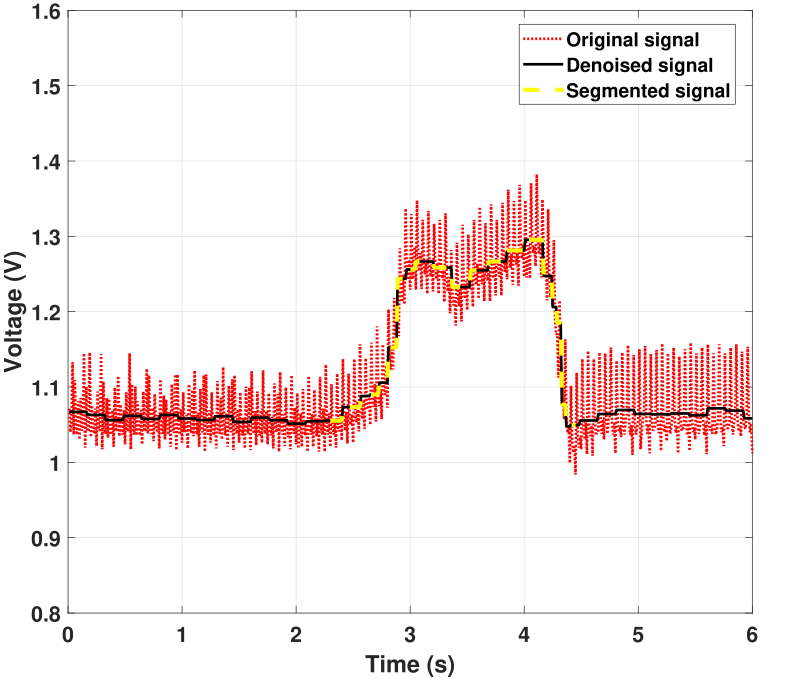}
\centering
\caption{The effect of Discrete Wavelet Denoising and Time Domain Thresholding blocks on the received signal at 20~cm.}
\label{fig:DWD_Simple_Thresholding_effect}
\vspace{-4mm}
\end{figure}

\begin{figure}[htbp]
\includegraphics[width=0.46\textwidth]{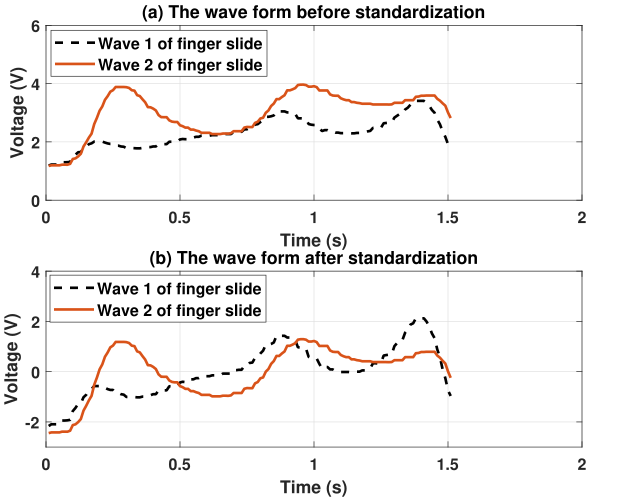}
\centering
\caption{The effect of Z Scores Standardization block on the time domain signal from received infrared light signal at 20~cm.}
\label{fig:standardization_effect}
\vspace{-4mm}
\end{figure}



\begin{figure*}[htbp]
\centering
\includegraphics[width=0.88\textwidth]{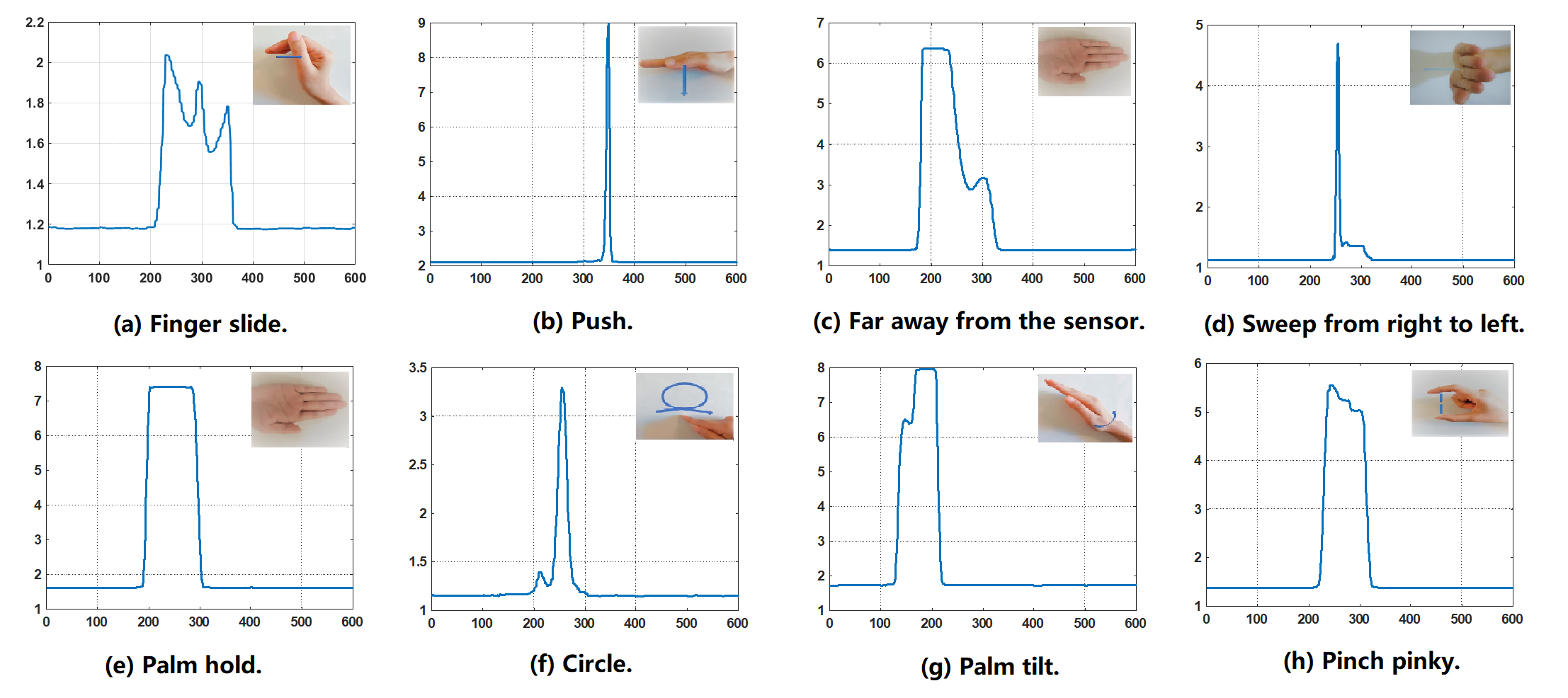}
\vspace{-3mm}
\caption{{Selected gesture classes with the corresponding time-series waves.}}
\label{fig:Gesture_set_1}
\vspace{-3mm}
\end{figure*}

Since gestures lasted for only 2-3 seconds, but data were collected for 6 seconds, hence most of the data at the beginning and at the end of the time-domain signal could be discarded to keep only the portion with higher intensity and larger fluctuation when the gesture was actually being performed. We used a simple thresholding method to segment the denoised data stream and mark the beginning and the end of the gesture. This thresholding was done in time domain, while the previous thresholding of wavelet co-efficients was done in frequency domain. The signals after denoising and time domain thresholding are shown with an example data in Fig.~\ref{fig:DWD_Simple_Thresholding_effect}. The segmented gesture signal had an unpredictable number of data points, hence zero-padding was done to make its length uniform for the gesture classification algorithm to work effectively later.

Then, Z scores were used to standardize  the signal amplitude based on the mean and variance of the data. Due to the diversity of factors involved when different users perform gestures, the magnitudes of measured light signals from the same gesture may be highly variable, even though their wave shapes are generally similar.  Z score is a standardization method that simply converts a data set to a distribution with zero mean and unity standard deviation \cite{zs}. The effectiveness of Z scores standardization on time domain signal is shown in Fig.~\ref{fig:standardization_effect}.  



\subsection{Classifier Training}      
After the signal processing was completed, all of the data sets were used as feature vectors in a training set to build a gesture classification method using the K-nearest-neighbors (KNN) algorithm~\cite{guo2003knn, peterson2009}. KNN is a non-parametric supervised machine learning algorithm that calculates distances between the test data and each training data, considers $k$ nearest training data and takes majority vote to decide which class the test data belongs to.

Hand gestures were selected {based on the common HCI tasks}\cite{hcixin1,light16}.  The waves shown in Fig.~\ref{fig:Gesture_set_1} are the {denoised}  time-domain signals for the selected gestures. After data analysis, we found that the variance of our feature vectors was small. This could also be observed visually from relatively small differences in the 8 gestures in the time-domain waves.  Therefore, due to the small difference and strong similarity of all the waves, KNN was found to be 
suitable for our method.

\section{Evaluation, Results and Discussion}
\label{sec:Results}
We evaluate our gesture recognition system using the collected data from real volunteers. First, we discuss the data collection method. 
Then, we describe a K-fold cross-validation procedure used to evaluate the gesture classification performance.
Further, we present an analysis of the gesture classification accuracy at different distances. Finally, the impact of environmental lighting conditions on gesture classification accuracy is presented.

 We instructed volunteers how to perform each gesture, and gave them several minutes to practice until they were comfortable with each gesture.  We recorded gesture data of 5 volunteers performing 24 repetitions of the 8 gestures (960 waves) shown in Fig.~\ref{fig:Gesture_set_1} at a distance of 20\,cm from the infrared sensor in normal indoor lighting conditions. To observe the impact of ambient light, the same number of waves were recorded at the same 20\,cm distance in the dark indoor (no lighting) condition too. The condition of ambient light is \emph{on} means that the fluorescent lights in the ceiling and the computer screen light are all included in the room lighting condition. And the condition of ambient light is \emph{off} means that the fluorescent lights in the ceiling are off and the computer screen is turned away from the photodetector making the effects of environment lighting zero.  Further, to obtain the effect of sensing distance on accuracy, the same number of waves were collected at distance 35\,cm from the sensor in normal indoor lighting conditions. Therefore, there were 3 data sets with the same number of waves in each.
 
\begin{table}[t]
\centering
\begin{tabular}{@{}clcccccccc@{}}
\toprule
\multicolumn{1}{l}{}                                                                                 & \multicolumn{9}{c}{Estimated Gesture}                                                                                                                                                                                                           \\ \midrule
\multicolumn{1}{|c|}{\multirow{9}{*}{\begin{tabular}[c]{@{}c@{}}\rotatebox{90}{Performed Gesture\hspace{1cm}}\end{tabular}}} & \multicolumn{1}{l|}{}    & \multicolumn{1}{l|}{(a)} & \multicolumn{1}{l|}{(b)} & \multicolumn{1}{l|}{(c)} & \multicolumn{1}{l|}{(d)} & \multicolumn{1}{l|}{(e)} & \multicolumn{1}{l|}{(f)} & \multicolumn{1}{l|}{(g)} & \multicolumn{1}{l|}{(h)} \\ \cmidrule(l){2-10} 
\multicolumn{1}{|c|}{}                                                                               & \multicolumn{1}{l|}{(a)} & \textbf{0.93}            & 0                        & 0.04                     & 0                        & 0                        & 0.01                     & 0                        & 0.01                     \\ \cmidrule(lr){2-2}
\multicolumn{1}{|c|}{}                                                                               & \multicolumn{1}{l|}{(b)} & 0             & \textbf{0.92}            & 0                        & 0.06                     & 0                        & 0.02                     & 0                        & 0                        \\ \cmidrule(lr){2-2}
\multicolumn{1}{|c|}{}                                                                               & \multicolumn{1}{l|}{(c)} & 0                        & 0                        & \textbf{1}               & 0                        & 0                        & 0                        & 0                        & 0                        \\ \cmidrule(lr){2-2}
\multicolumn{1}{|c|}{}                                                                               & \multicolumn{1}{l|}{(d)} & 0.01                     & 0.01                     & 0                        & \textbf{0.97}            & 0                        & 0.01                     & 0                        & 0                        \\ \cmidrule(lr){2-2}
\multicolumn{1}{|c|}{}                                                                               & \multicolumn{1}{l|}{(e)} & 0                        & 0                        & 0                        & 0                        & \textbf{0.98}            & 0                        & 0.02                     & 0                        \\ \cmidrule(lr){2-2}
\multicolumn{1}{|c|}{}                                                                               & \multicolumn{1}{l|}{(f)} & 0.01                     & 0                        & 0                        & 0.03                     & 0                        & \textbf{0.96}            & 0                        & 0                        \\ \cmidrule(lr){2-2}
\multicolumn{1}{|c|}{}                                                                               & \multicolumn{1}{l|}{(g)} & 0                        & 0                        & 0.03                     & 0                        & 0                        & 0                        & \textbf{0.97}            & 0                        \\ \cmidrule(lr){2-2}
\multicolumn{1}{|c|}{}                                                                               & \multicolumn{1}{l|}{(g)}                      & 0.02                     & 0                        & 0                        & 0.01                     & 0                        & 0.01                     & 0                        & \textbf{0.96}            \\ \bottomrule
\end{tabular}
\caption{The 10-fold cross-validation confusion matrix of infrared light wave sensing around 20\,cm (ambient light is on).}
\label{inf1}
\vspace{-2mm}
\end{table}

Next, we evaluated the accuracy of our system at different distances, and in different ambient lighting conditions. All the results were obtained with 10-fold cross-validation.  In 10-fold cross-validation, all the waves were divided into 10 subsets of equal size randomly. One of the 10 subsets was taken as testing data, the remaining 9 subsets were used as training data. Then the cross-validation was repeated 10 times, and each of the 10 subsets was used as the testing data only once. Therefore, 10 confusion matrices were averaged to obtain a single result. The confusion matrix for infrared sensing around 20\,cm are shown in Table~\ref{inf1}. The overall accuracy of found through cross-validation considering all gestures was 96.13\% (SD = 2.59\%).

\subsubsection{Accuracy at different distances}       

The effect of distance on gesture recognition accuracy is shown in Fig.~\ref{fig:infrared_comparison}. Accuracy decreased with increasing distance because of the lower reflected light intensity that was reducing the overall signal to noise ratio (SNR) of the system.

\begin{table}[t]
\centering
\begin{tabular}{@{}clcccccccc@{}}
\toprule
\multicolumn{1}{l}{}                                                                                 & \multicolumn{9}{c}{Estimated Gesture}                                                                                                                                                                                                           \\ \midrule
\multicolumn{1}{|c|}{\multirow{9}{*}{\begin{tabular}[c]{@{}c@{}}\rotatebox{90}{Performed Gesture\hspace{1cm}}\end{tabular}}} & \multicolumn{1}{l|}{}    & \multicolumn{1}{l|}{(a)} & \multicolumn{1}{l|}{(b)} & \multicolumn{1}{l|}{(c)} & \multicolumn{1}{l|}{(d)} & \multicolumn{1}{l|}{(e)} & \multicolumn{1}{l|}{(f)} & \multicolumn{1}{l|}{(g)} & \multicolumn{1}{l|}{(h)} \\ \cmidrule(l){2-10} 
\multicolumn{1}{|c|}{}                                                                               & \multicolumn{1}{l|}{(a)} &   \textbf{0.94}&0&0&0&0&0&0&0.06               \\ \cmidrule(lr){2-2}
\multicolumn{1}{|c|}{}                                                                               & \multicolumn{1}{l|}{(b)} &    0& \textbf{0.97}&0&0&0&0.03&0&0                     \\ \cmidrule(lr){2-2}
\multicolumn{1}{|c|}{}                                                                               & \multicolumn{1}{l|}{(c)} &   0 &0& \textbf{0.99}&0&0&0&0.01&0                        \\ \cmidrule(lr){2-2}
\multicolumn{1}{|c|}{}                                                                               & \multicolumn{1}{l|}{(d)} &    0&0.02&0& \textbf{0.98}&0&0&0&0                 \\ \cmidrule(lr){2-2}
\multicolumn{1}{|c|}{}                                                                               & \multicolumn{1}{l|}{(e)} &    0 &0&0&0& \textbf{0.98}&0&0.02&0           \\ \cmidrule(lr){2-2}
\multicolumn{1}{|c|}{}                                                                               & \multicolumn{1}{l|}{(f)} &    0.03 &0.02&0&0&0& \textbf{0.92}&0&0.03                 \\ \cmidrule(lr){2-2}
\multicolumn{1}{|c|}{}                                                                               & \multicolumn{1}{l|}{(g)} &    0&0&0.03&0&0&0& \textbf{0.97}&0                    \\ \cmidrule(lr){2-2}
\multicolumn{1}{|c|}{}                                                                               & \multicolumn{1}{l|}{(h)} &    0 &0&0&0&0&0&0& \textbf{1.00}      \\ \bottomrule
\end{tabular}
 \caption{The 10-fold cross-validation confusion matrix of infrared light wave sensing around 20\,cm in the dark (ambient light is off).}  
 \label{infdark}
\vspace*{-4mm}
\end{table}

\subsubsection{Accuracy with different ambient lighting}   
\begin{figure}[htbp]
\vspace{-4mm}
\includegraphics[width=0.46\textwidth]{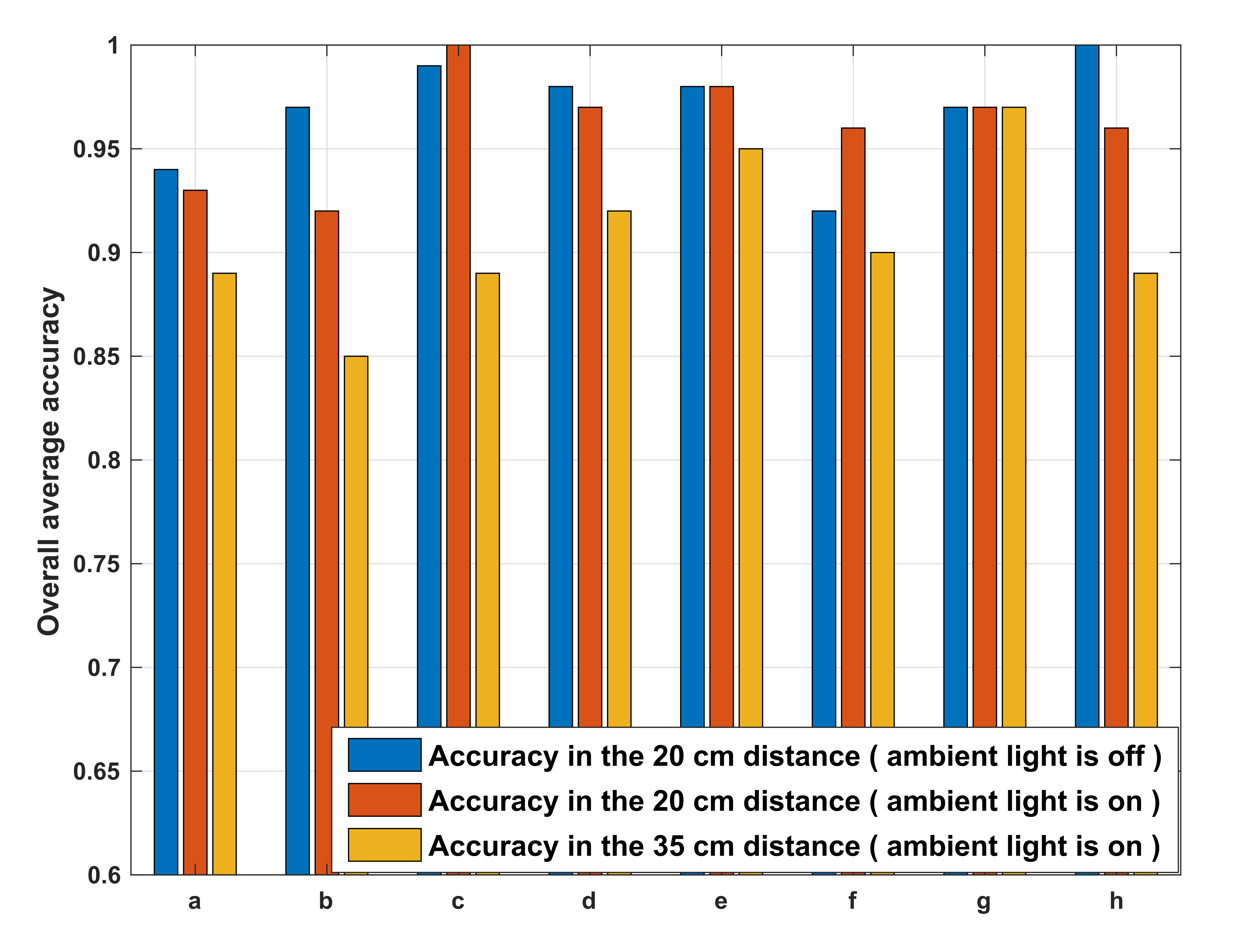}
\centering
\vspace{-4mm}
\caption{Gesture recognition accuracies at different distances and ambient lighting conditions.}
\label{fig:infrared_comparison}
\vspace{-4mm}
\end{figure}

To determine the effect of ambient light on recognition accuracy, we compared the classification results with ambient lights \emph{on} and \emph{off} at $d=20$\,cm. 
Ambient lighting conditions had little, if any, significant impact on the system accuracy, as seen from the results presented in Table~\ref{inf1},~\ref{infdark} and Fig.~\ref{fig:infrared_comparison}. This was attributed to the peak responsivity of the photodetector near infrared wavelength which caused the power of the reflected infrared light be much greater than the contribution from ambience~\cite{detector2}.

\section{Conclusion}
\label{sec:Conclusion}

We presented a non-contact gesture recognition system that utilized the variation of reflected incoherent infrared light intensity due to hand gestures performed within 20\,cm to 35\,cm distance from the light source in both \emph{on} and \emph{off} conditions of ambient lighting. This light-based technology is simple, low-cost, safe and portable in nature. Infrared light signals can be easily generated and analysed too. Moreover, we employed a series of signal processing and machine learning algorithms in the system such that the sensing modality could achieve high accuracy in recognizing 8 gestures. Future studies will be conducted to better quantify and model the operation of the system, to further verify this method's practicality and limitations, and to improve the system's performance.



\bibliographystyle{IEEEtran}
\bibliography{ref.bib}

\begin{thebibliography}{10}
\providecommand{\url}[1]{#1}
\csname url@samestyle\endcsname
\providecommand{\newblock}{\relax}
\providecommand{\bibinfo}[2]{#2}
\providecommand{\BIBentrySTDinterwordspacing}{\spaceskip=0pt\relax}
\providecommand{\BIBentryALTinterwordstretchfactor}{4}
\providecommand{\BIBentryALTinterwordspacing}{\spaceskip=\fontdimen2\font plus
\BIBentryALTinterwordstretchfactor\fontdimen3\font minus
  \fontdimen4\font\relax}
\providecommand{\BIBforeignlanguage}[2]{{%
\expandafter\ifx\csname l@#1\endcsname\relax
\typeout{** WARNING: IEEEtran.bst: No hyphenation pattern has been}%
\typeout{** loaded for the language `#1'. Using the pattern for}%
\typeout{** the default language instead.}%
\else
\language=\csname l@#1\endcsname
\fi
#2}}
\providecommand{\BIBdecl}{\relax}
\BIBdecl

\bibitem{IOT1}
\BIBentryALTinterwordspacing
J.~Gubbi, R.~Buyya, S.~Marusic, and M.~Palaniswami, ``{Internet of Things
  (IoT): A vision, architectural elements, and future directions},''
  \emph{Future Generation Computer Systems}, vol.~29, no.~7, pp. 1645--1660,
  2013. [Online]. Available:
  \url{https://www.sciencedirect.com/science/article/pii/S0167739X13000241}
\BIBentrySTDinterwordspacing

\bibitem{hcixin}
J.~Lazar, J.~H. Feng, and H.~Hochheiser, \emph{Research methods in
  human-computer interaction}.\hskip 1em plus 0.5em minus 0.4em\relax Morgan
  Kaufmann, 2017.

\bibitem{wear1}
C.~Zhu and W.~Sheng, ``{Wearable Sensor-Based Hand Gesture and Daily Activity
  Recognition for Robot-Assisted Living},'' \emph{IEEE Transactions on Systems,
  Man, and Cybernetics - Part A: Systems and Humans}, vol.~41, no.~3, pp.
  569--573, May 2011.

\bibitem{wear4}
M.~Caputo, K.~Denker, B.~Dums, G.~Umlauf, H.~Konstanz, and G.~, ``{3D Hand
  Gesture Recognition Based on Sensor Fusion of Commodity Hardware},'' vol.
  2012, 01 2012.

\bibitem{RF1}
\BIBentryALTinterwordspacing
F.~Adib, Z.~Kabelac, D.~Katabi, and R.~C. Miller, ``{3D Tracking via Body Radio
  Reflections},'' in \emph{11th {USENIX} Symposium on Networked Systems Design
  and Implementation ({NSDI} 14)}.\hskip 1em plus 0.5em minus 0.4em\relax
  Seattle, WA: {USENIX} Association, 2014, pp. 317--329. [Online]. Available:
  \url{https://www.usenix.org/conference/nsdi14/technical-sessions/presentation/adib}
\BIBentrySTDinterwordspacing

\bibitem{RF2}
\BIBentryALTinterwordspacing
Y.~Zeng, P.~H. Pathak, and P.~Mohapatra, ``{Analyzing Shopper's Behavior
  Through WiFi Signals},'' in \emph{Proceedings of the 2Nd Workshop on Workshop
  on Physical Analytics}, ser. WPA '15.\hskip 1em plus 0.5em minus 0.4em\relax
  New York, NY, USA: ACM, 2015, pp. 13--18. [Online]. Available:
  \url{http://doi.acm.org/10.1145/2753497.2753508}
\BIBentrySTDinterwordspacing

\bibitem{RF3}
\BIBentryALTinterwordspacing
S.~Sen, J.~Lee, K.-H. Kim, and P.~Congdon, ``Avoiding multipath to revive
  inbuilding wifi localization,'' in \emph{Proceeding of the 11th Annual
  International Conference on Mobile Systems, Applications, and Services}, ser.
  MobiSys '13.\hskip 1em plus 0.5em minus 0.4em\relax New York, NY, USA: ACM,
  2013, pp. 249--262. [Online]. Available:
  \url{http://doi.acm.org/10.1145/2462456.2464463}
\BIBentrySTDinterwordspacing

\bibitem{RF4}
\BIBentryALTinterwordspacing
J.~Lien, N.~Gillian, M.~E. Karagozler, P.~Amihood, C.~Schwesig, E.~Olson,
  H.~Raja, and I.~Poupyrev, ``{Soli: Ubiquitous Gesture Sensing with Millimeter
  Wave Radar},'' \emph{ACM Transactions on Graphics}, vol.~35, no.~4, pp.
  1--19, Jul. 2016. [Online]. Available:
  \url{https://dl.acm.org/doi/10.1145/2897824.2925953}
\BIBentrySTDinterwordspacing

\bibitem{image1}
L.~Chen, H.~Wei, and J.~Ferryman, ``A survey of human motion analysis using
  depth imagery,'' \emph{Pattern Recognition Letters}, vol.~34, no.~15, pp.
  1995 -- 2006, 2013, smart Approaches for Human Action Recognition.

\bibitem{image2}
\BIBentryALTinterwordspacing
M.~A.~R. Ahad, J.~K. Tan, H.~Kim, and S.~Ishikawa, ``Motion history image: Its
  variants and applications,'' \emph{Mach. Vision Appl.}, vol.~23, no.~2, p.
  255–281, Mar. 2012. [Online]. Available:
  \url{https://doi.org/10.1007/s00138-010-0298-4}
\BIBentrySTDinterwordspacing

\bibitem{light7}
\BIBentryALTinterwordspacing
R.~H. Venkatnarayan and M.~Shahzad, ``{Gesture Recognition Using Ambient
  Light},'' \emph{Proceedings of the ACM on Interactive, Mobile, Wearable and
  Ubiquitous Technologies}, vol.~2, no.~1, pp. 1--28, Mar. 2018. [Online].
  Available: \url{https://dl.acm.org/doi/10.1145/3191772}
\BIBentrySTDinterwordspacing

\bibitem{light16}
M.~Kaholokula, ``{Reusing Ambient Light to Recognize Hand Gestures},''
  Dartmouth College, Tech. Rep., 2016.

\bibitem{soundnew1}
Y.~{Qifan}, T.~{Hao}, Z.~{Xuebing}, L.~{Yin}, and Z.~{Sanfeng}, ``Dolphin:
  Ultrasonic-based gesture recognition on smartphone platform,'' in \emph{2014
  IEEE 17th International Conference on Computational Science and Engineering},
  2014, pp. 1461--1468.

\bibitem{soundnew2}
\BIBentryALTinterwordspacing
A.~Mujibiya, X.~Cao, D.~S. Tan, D.~Morris, S.~N. Patel, and J.~Rekimoto, ``The
  sound of touch: On-body touch and gesture sensing based on transdermal
  ultrasound propagation,'' in \emph{Proceedings of the 2013 ACM International
  Conference on Interactive Tabletops and Surfaces}, ser. ITS ’13.\hskip 1em
  plus 0.5em minus 0.4em\relax NY, USA: Assoc. for Computing Machinery, 2013,
  p. 189–198. [Online]. Available:
  \url{https://doi.org/10.1145/2512349.2512821}
\BIBentrySTDinterwordspacing

\bibitem{emirf1}
Z.~Chi, Y.~Yao, T.~Xie, X.~Liu, Z.~Huang, W.~Wang, and T.~Zhu, ``{EAR:
  Exploiting uncontrollable ambient RF signals in heterogeneous networks for
  gesture recognition},'' in \emph{{SenSys 2018 - Proceedings of the 16th
  Conference on Embedded Networked Sensor Systems}}, Nov 2018, pp. 237--249.

\bibitem{emirf2}
Z.~Tian, X.~Yang, and M.~Zhou, ``{WiCatch: A Wi-Fi Based Hand Gesture
  Recognition System},'' \emph{IEEE Access}, vol.~6, pp. 16\,911--16\,923, Mar
  2018.

\bibitem{geisheimer2001non}
J.~Geisheimer and E.~Greneker, ``A non-contact lie detector using radar vital
  signs monitor (rvsm) technology,'' \emph{IEEE Aerospace and Electronic
  Systems Magazine}, vol.~16, no.~8, pp. 10--14, 2001.

\bibitem{masao2001biological}
T.~Masao and S.~WATANABE, ``Biological and health effects of exposure to
  electromagnetic field from mobile communications systems,'' \emph{IATSS
  research}, vol.~25, no.~2, pp. 40--50, 2001.

\bibitem{camera5}
S.~Oprisescu, C.~Rasche, and B.~Su, ``Automatic static hand gesture recognition
  using tof cameras,'' in \emph{2012 Proceedings of the 20th European Signal
  Processing Conference (EUSIPCO)}, Aug 2012, pp. 2748--2751.

\bibitem{camera6}
T.~Plotz, C.~Chen, N.~Y. Hammerla, and G.~D. Abowd, ``Automatic synchronization
  of wearable sensors and video-cameras for ground truth annotation -- a
  practical approach,'' in \emph{2012 16th International Symposium on Wearable
  Computers}, June 2012, pp. 100--103.

\bibitem{soundnew}
H.~Watanabe, T.~Terada, and M.~Tsukamoto, ``{Ultrasound-Based Movement Sensing,
  Gesture-, and Context-Recognition},'' in \emph{Proceedings of the 2013
  International Symposium on Wearable Computers}, ser. ISWC ’13.\hskip 1em
  plus 0.5em minus 0.4em\relax New York, NY, USA: Association for Computing
  Machinery, 2013, p. 57–64.

\bibitem{transmitter}
\BIBentryALTinterwordspacing
{940nm IR lamp Board with Light Sensor (48 Black LED Illuminator Array)}.
  Amazon.com, Inc. Accessed on: 07-11-2020. [Online]. Available:
  \url{https://www.amazon.com/gp/product/B0785W2RQQ}
\BIBentrySTDinterwordspacing

\bibitem{detector}
\BIBentryALTinterwordspacing
{PDA100A}. Thorlabs, Inc. Accessed on: 07-11-2020. [Online]. Available:
  \url{https://www.thorlabs.com/thorproduct.cfm?partnumber=PDA100A}
\BIBentrySTDinterwordspacing

\bibitem{dwt}
J.~E. {Fowler}, ``{The redundant discrete wavelet transform and additive
  noise},'' \emph{IEEE Signal Processing Letters}, vol.~12, no.~9, pp.
  629--632, Sep. 2005.

\bibitem{dwtxin}
\BIBentryALTinterwordspacing
E.~Alickovic, J.~Kevric, and A.~Subasi, ``Performance evaluation of empirical
  mode decomposition, discrete wavelet transform, and wavelet packed
  decomposition for automated epileptic seizure detection and prediction,''
  \emph{Biomedical Signal Processing and Control}, vol.~39, pp. 94--102, 2018.
  [Online]. Available:
  \url{https://www.sciencedirect.com/science/article/pii/S1746809417301544}
\BIBentrySTDinterwordspacing

\bibitem{zs}
\BIBentryALTinterwordspacing
N.~Pollesch and V.~Dale, ``{Normalization in sustainability assessment: Methods
  and implications},'' \emph{Ecological Economics}, vol. 130, pp. 195--208,
  2016. [Online]. Available:
  \url{https://www.sciencedirect.com/science/article/pii/S0921800915305899}
\BIBentrySTDinterwordspacing

\bibitem{guo2003knn}
G.~Guo, H.~Wang, D.~Bell, Y.~Bi, and K.~Greer, ``{KNN model-based approach in
  classification},'' in \emph{OTM Confederated International Conferences" On
  the Move to Meaningful Internet Systems"}.\hskip 1em plus 0.5em minus
  0.4em\relax Springer, 2003, pp. 986--996.

\bibitem{peterson2009}
L.~E. Peterson, ``K-nearest neighbor,'' \emph{Scholarpedia}, vol.~4, no.~2, p.
  1883, 2009.

\bibitem{hcixin1}
\BIBentryALTinterwordspacing
S.~Amershi, M.~Cakmak, W.~B. Knox, and T.~Kulesza, ``Power to the people: The
  role of humans in interactive machine learning,'' \emph{AI Magazine},
  vol.~35, no.~4, pp. 105--120, Dec. 2014. [Online]. Available:
  \url{https://www.aaai.org/ojs/index.php/aimagazine/article/view/2513}
\BIBentrySTDinterwordspacing

\bibitem{detector2}
\BIBentryALTinterwordspacing
{Si Free-space Gain Detector User Guide}. Thorlabs, Inc. Accessed on:
  07-05-2022. [Online]. Available:
  \url{https://www.thorlabs.com/catalogpages/Obsolete/2018/PDA100A.pdf}
\BIBentrySTDinterwordspacing

\end{thebibliography}


\vspace{12pt}
\color{red}

\end{document}